\documentclass[12pt]{article}
\usepackage{amsmath}
\usepackage{graphics}

\begin{document}
\begin{center}
{\LARGE Are atoms waves or particles?}\\
Trevor W. Marshall\\
\emph {Dept. of Mathematics, Univ. of Manchester, Manchester M13
9PL, UK}
\end{center}
{\bf Abstract}

It is shown that the Kapitza-Dirac effect with atoms, which has
been considered to be evidence for their wavelike character, can
be interpreted as a scattering of pointlike objects by the
periodic laser field.
\section{Introduction}

The currently accepted answer to the question posed in my title
is, of course, "Both". But I submit that we should not abandon the
heritage passed down to us by the Atomists, from Democritus to
Boltzmann. It was a long struggle, at times involving scientific
isolation and consequent personal suffering\cite{brush}, to
establish the atomicity of matter (from which I exclude radiation
for present purposes).

In recent times some \ experimental evidence has been found to
support a wavelike description of atoms, and even of quite large
molecules such as fullerene. I shall concentrate here on the first
category, in which something analogous to the diffraction of light
has been observed with a "monochromatic" beam of atoms, the
grating being supplied by a stationary
laser wave which is tuned to a frequency close to an atomic resonance\cite%
{gould}. Actually it is not so much the monochrome property which
is important -- the velocity of the beam was controlled only to
within about 5\% of its mean value -- but rather a very high
degree of collimation -- the component of momentum perpendicular
to the laser must be an order of
magnitude less than $h/\lambda $, where $h$ is Planck's constant and $%
\lambda $ is the laser wave length. For sodium atoms with a mean
velocity of 10$^{3}$ms$^{-1}$, and with the laser tuned to the
D-line at 589nm, this demands an initial angular spread less than
3.10$^{-5}$radians. What we observe in the outgoing beam is a set
of well separated peaks at integral multiples of 6.10$^{-5}$rad.
There is a well worked out theory of the broadening of these lines\cite%
{gould1}\cite{gould2}, but the spacing of $2h/\lambda Mv$, as well
as the intensities, may be explained with a very simple quantum
mechanical (QM) model to be outlined in the next section. This
model was discussed by Gould\cite{gould}, who offered two
interpretations of the analysis; we must accept \emph{either }that
each atom
is spread out over a wave front of the order of several microns \emph{or }%
that the atom trades in quanta of momentum. The first alternative
is the
description given long ago by Kapitza and Dirac\cite{kapitza} of a \emph{%
diffraction} process in which an atomic wave whose wavelength is
$h/Mv$ is diffracted by a grating whose spacing is $\lambda /2$,
while the second views the process as one of \emph{scattering }in
which the atom absorbs and
emits, stochastically, radiation from and to the laser field in quanta of $%
hc/\lambda $; such radiation is in one of the two (up or down)
directions of
the laser beam, and so its Poynting vector carries a transverse momentum of $%
\pm h/\lambda $, and this must be compared with a longitudinal momentum of $%
Mv.$ But, curiously, the latter analysis indicates that events of
emission and absorption occur in pairs, so that very few atoms
emerge from the laser having a transverse momentum which is an odd
multiple of $h/\lambda $.

Gould did not choose to emphasize the contradictory nature of
these two interpretations; he instead pointed out that either of
them were "equally unpalatable to the prequantum physicist".
Staying within the Atomist tradition I propose to reject the first
interpretation and accept the second. Nevertheless, I enter the
reservation that I can do so staying largely within a classical
(or prequantum) world view. There is a fair amount of
evidence\cite{kuhn} that Max Planck, who discovered the quantum
discontinuity in absorption and emission of light, never accepted
that the light field itself had to be quantized. A quotation from
a letter to Einstein in 1907 illustrates Planck's view of the
light field\emph{.}

\begin{quotation}
I am not seeking the meaning of the quantum of action (light
quantum) in the vacuum but rather in places where emission and
absorption occur, and I assume that what happens in the vacuum is
rigorously described by Maxwell's equations.
\end{quotation}

In summary, I propose, from Section 3 onwards, to investigate
whether the distribution pattern of the scattered atoms may be
explained on the basis of a model in which quanta $h/\lambda $ of
momentum are exchanged, stochastically, with the laser field.
Before that I shall summarize, in Section 2, the results of the
simplified QM theory, which will provide us with a standard for
comparison.
\section{The QM model}

The hamiltonian is
\begin{equation}
H\left( t\right) =\frac{1}{2}\hbar \left( \omega -\Delta \right)
\sigma _{3}+\hbar \Omega _{R}\cos \zeta \left( \sigma _{1}\cos
\omega t+\sigma _{2}\sin \omega t\right) \text{ },\text{
}0<t<t_{0}\text{ ,}
\end{equation}%
where $\omega $ is the laser frequency, detuned by $\Delta $ from
the D-line resonance $\omega_0$, $\Omega _{R}$ is the resonant
Rabi frequency of the interaction, and $\sigma _{1},\sigma
_{2},\sigma _{3}$ are the Pauli spin matrices interpreted, in the
standard manner, as atomic raising and lowering operators. The
kinetic energy $\left( h^{2}/2M\lambda ^{2}\right) \left(
\partial ^{2}/\partial \zeta ^{2}\right) $ has been discarded on account of
the large atomic mass $M,$ and the variable $\zeta ,$ which takes
values in the range $\left( -\pi /2,\pi /2\right) $ (see Fig.1),
\begin{figure}[htb]
\begin{center}
\scalebox{0.5}{\includegraphics{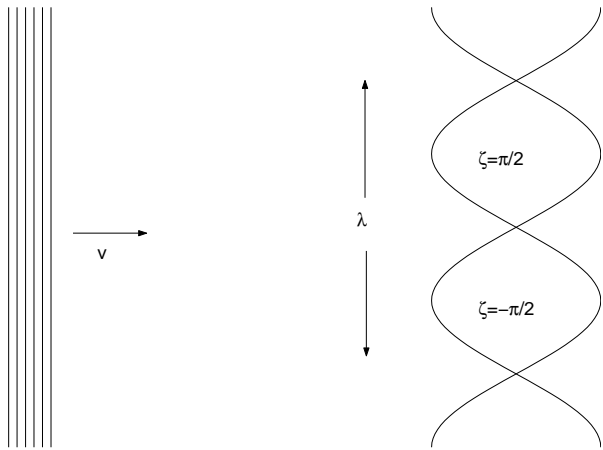}}
\end{center}
\caption{The Kapitza-Dirac effect according to the QM model, which
is the original description of Dirac and Kapitza. A de Broglie
wave, whose
wavelength is small, and whose coherence width is large, compared with $%
\protect\lambda $, is incident on a stationary laser of wavelength $\protect%
\lambda $. The de Broglie wave experiences the laser as a
sinusoidally varying refractive index with spacing $\lambda/2$,
and emerges as a diffraction pattern whose maxima are separated by
angular intervals of $2h/\protect\lambda Mv$ radians.}
\end{figure}
gives the phase of the atom in the laser wave at the point of
entry. On account of the assumption of infinite mass, this is also
the phase at the point of exit. Starting from an initial state
having zero transverse momentum and in the lower state of the
D-line couplet, that is%
\begin{equation}
\psi \left( 0;\zeta \right) =\left(
\begin{array}{c}
0 \\
1%
\end{array}%
\right) \text{ ,}
\end{equation}%
the state at time $t_{0}$ is%
\begin{equation}
\left(
\begin{array}{c}
\psi _{u}\left( t_{0};\zeta \right) e^{i\omega t_{0}/2} \\
\psi _{l}\left( t_{0};\zeta \right) e^{-i\omega t_{0}/2}%
\end{array}%
\right) =\left(
\begin{array}{c}
\left[ -i(\Omega _{R}/\Omega )\cos \zeta \sin \Omega t_{0}\right]
e^{i\omega
t_{0}/2} \\
\left[ \cos \Omega t_{0}+i\left( \Delta /2\Omega \right) \sin \Omega t_{0}%
\right] e^{-i\omega t_{0}/2}%
\end{array}%
\right) \text{ ,}  \label{wavefun}
\end{equation}%
where $\Omega =\sqrt{\left( \Delta ^{2}/4\right) +\Omega
_{R}^{2}\cos ^{2}\zeta }$. The amplitudes $\psi _{u}$ and $\psi
_{l}$ are the parts of the wave function representing an atom in
its upper and lower state respectively. In order to observe the
discrete momentum spectrum it was found necessary to make $\Delta
$ substantially larger than $\Omega _{R};$ typically $\Delta
\approx 5\Omega _{R}$. We therefore expand the wave function in
powers of $\gamma =\Omega _{R}/\Delta $ giving, to order $\gamma
^{2}$,%
\begin{eqnarray}
\psi _{u}\left( t_{0};\zeta \right) &=&-2i\gamma \cos \zeta \sin
\Omega
^{\prime }\tau \text{ ,}  \notag \\
\psi _{l}\left( t_{0};\zeta \right) &=&e^{i\Omega ^{\prime }\tau
}-2i\gamma ^{2}\cos ^{2}\zeta \sin \Omega ^{\prime }\tau \text{ ,}
\end{eqnarray}%
where%
\begin{equation}
\tau =t_{0}\gamma ^{2}\Delta /2\qquad ,\qquad \Omega ^{\prime
}=\gamma ^{-2}+1+\cos 2\zeta \text{ .}
\end{equation}%
The Fourier series for the transition amplitude is then%
\begin{equation}
\psi _{u}\left( \tau ;\zeta \right) =-i\gamma \sum_{n=-\infty
}^{\infty } \left[ J_{n}\left( \tau \right) \sin \tau
_{n}+J_{n+1}\left( \tau \right) \cos \tau _{n}\right] e^{\left(
2n+1\right) i\zeta }\text{ ,}
\end{equation}%
where%
\begin{equation}
\tau _{n}=\frac{\left( 1+\gamma ^{2}\right) \tau }{\gamma ^{2}}+\frac{n\pi }{%
2}\text{ ,}  \label{taun}
\end{equation}%
and for the no-transition amplitude it is%
\begin{equation}
\psi _{l}\left( \tau ;\zeta \right) =\sum_{-\infty }^{\infty
}e^{2in\zeta } \left[ e^{i\tau _{n}}J_{n}\left( \tau \right)
-i\gamma ^{2}\{J_{n}\left( \tau \right) \sin \tau
_{n}-J_{n}^{\prime }\left( \tau \right) \cos \tau _{n}\}\right]
\text{ .}
\end{equation}%
The lower component gives the intensities of the even lines of the
spectrum,
namely%
\begin{equation}
\rho _{n}^{\text{Q}}\left( \tau \right) =J_{n}^{2}\left( \tau
\right) \left( 1-2\gamma ^{2}\sin ^{2}\tau _{n}\right) +\gamma
^{2}J_{n}\left( \tau \right) J_{n}^{\prime }\left( \tau \right)
\sin 2\tau _{n}\text{ ,} \label{qspeceven}
\end{equation}%
while the upper component gives the odd lines, namely%
\begin{equation}
\rho _{n+1/2}^{\text{Q}}\left( \tau \right) =\gamma ^{2}\left[
J_{n}\left( \tau \right) \sin \tau _{n}+J_{n+1}\left( \tau \right)
\cos \tau _{n}\right] ^{2}\text{ .}  \label{qspecodd}
\end{equation}
The wave interpretation of Kapitza and Dirac is made by
considering the limit $\gamma \rightarrow 0$, so that the outgoing
wave function is effectively the
scalar%
\begin{equation}
\psi \left( t_{0};\zeta \right) =\exp \left[ i\left(
-\frac{1}{2}\omega _{0}t_{0}+2\tau \cos ^{2}\zeta \right) \right]
\text{ .}
\end{equation}%
The dependence of $\psi $ on $\zeta $ is an indication (see Fig.1)
that the de Broglie wave of the atom experiences a spatially
varying refractive index
as it goes through the laser, and its Fourier expansion, that is%
\begin{equation}
\exp \left( 2i\tau \cos ^{2}\zeta \right) =\sum_{n=-\infty
}^{\infty }i^{n}e^{i\tau }J_{n}\left( \tau \right) e^{2in\zeta
}\text{ ,}
\end{equation}%
indicates that the atom acquires a transverse momentum of either $%
2nh/\lambda $ or $-2nh/\lambda $ with probability%
\begin{equation}
\rho _{n}^{\text{Q}0}\left( \tau \right) =J_{n}^{2}\left( \tau
\right) \text{ ,}
\end{equation}%
for which the characteristic function is%
\begin{equation}
F_{\text{Q}0}\left( \theta ;\tau \right) =\left\langle e^{in\theta
}\right\rangle =\sum_{n=-\infty }^{\infty }J_{n}^{2}\left( \tau
\right) e^{in\theta }=J_{0}\left( 2\tau \sin \frac{\theta
}{2}\right) \text{ .}
\end{equation}%

The moments of the distribution are obtained from the derivatives
of $F$. In
particular the variance is%
\begin{equation}
\left\langle n^{2}\right\rangle _{\text{Q}0}=\sum_{n=-\infty
}^{\infty }n^{2}J_{n}^{2}\left( \tau \right)
=-F_{\text{Q}0}^{\prime \prime }\left( 0;\tau \right)
=\frac{1}{2}\tau ^{2}\text{ .}
\end{equation}%
This has the form, for small $\tau ,$%
\begin{equation}
\left\langle n^{2}\right\rangle _{\text{Q}0}=2\rho
_{1}^{\text{Q}0}\left( \tau \right) +O\left( \tau ^{4}\right)
\text{ ,}  \label{sstp}
\end{equation}%
which establishes that changes in $n$ occur in single steps of
$\pm 1$.

We shall need to consider the asymptotic behaviour as
$\tau\rightarrow\infty$
of this spectrum, namely%
\cite{GR}
\footnote{%
In the transition region $n\approx\tau$ these asymptotic
expressions should be replaced by others, obtained from Airy
approximations and also given in \cite{GR}. However, the velocity
averaging which I propose next will mask this correction.}
\begin{equation}
\rho _{n}^{\text{Q}0}\left( \tau \right) \sim \left\{
\begin{array}{c}
2\pi ^{-1}\left( \tau ^{2}-n^{2}\right) ^{-1/2}\cos ^{2}\left[
\sqrt{\tau ^{2}-n^{2}}-\beta |n|-\pi /4\right] \quad \left(
|n|<\tau \right) \text{ ,}
\\
0\qquad \qquad \qquad \qquad \qquad \qquad \left( |n|>\tau \right) \text{ ,}%
\end{array}%
\right.
\end{equation}%
where%
\begin{equation}
\beta =\cos ^{-1}\left( |n|/\tau \right) \text{ .}
\end{equation}%
These intensities oscillate, with angular frequency $1/2$ for
small $n$, but decreasing as $n$ approaches $\tau $.
\begin{figure}[bth]
\begin{center}
\scalebox{0.5}{\includegraphics{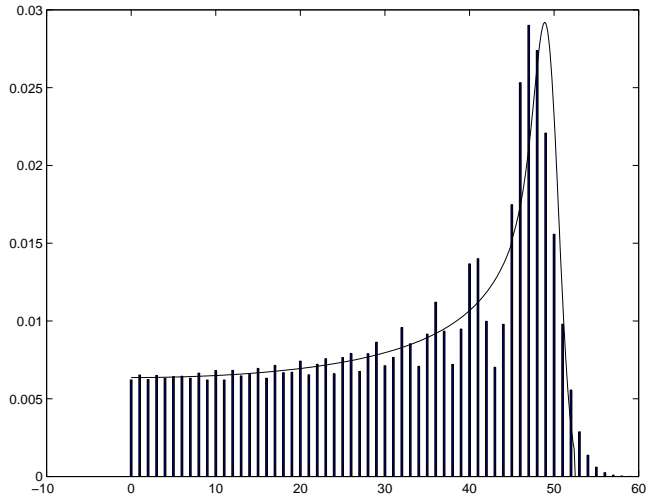}}
\end{center}
\caption{The  momentum spectrum, averaged over the velocity
profile, of the Kapitza-Dirac effect according to the QM model,
with the time parameter $\protect\tau=50$. The bar chart
represents the intensity of the $n$th line and the continuous
curve depicts a deterministic classical model. Since the observed
datum is actually the angular deflection, the experimental method
used cannot distinguish the two spectra at this value of $\tau$. }
\end{figure}
However, the oscillations disappear once we take account of the
beam's velocity profile, described by a gaussian function
$H(\tau)$, with standard deviation $\sigma =$0.025$\tau $, so that
95\% of the atoms have transit times within 5\% of the mean. The
averaged intensities are
\begin{equation}
\overline{\rho _{n}^{\text{Q}0}}\left( \tau \right)
=\int_{0}^{\infty }J_{n}^{2}\left( \tau ^{\prime }\right) H\left(
\tau ,\tau ^{\prime }\right)
d\tau ^{\prime }\sim \frac{1}{\pi \sigma \sqrt{2\pi }}\int_{n}^{\infty }\exp %
\left[ -\frac{\left( \tau -\tau ^{\prime }\right) ^{2}}{2\sigma
^{2}}\right] \frac{d\tau ^{\prime }}{\sqrt{\tau ^{\prime
2}-n^{2}}}\text{ ,} \label{hsmooth}
\end{equation}%
I have plotted, in Fig.2, the values of $\overline{\rho
_{n}^{\text{Q}0}}$ and its asymptotic limit at $\tau =50$. The
latter curve actually coincides with a completely classical model
of the process, in which an atom going through
the laser at phase $\zeta $ acquires a momentum of $\tau \sin 2\zeta ,$ and $%
\zeta $ has uniform density between $-\pi /2$ and $\pi /2$. For
such large $\tau $ it is this classical curve which would be
observed, because the experimental datum is the angular
deflection, rather than the transverse momentum, of the atom; the
individual lines of the spectrum are broadened, so that they merge
with one another.

The exact spectrum, given by (\ref{qspeceven}) and
(\ref{qspecodd}), displays rapid oscillations because of the
sinusoidal terms in $\tau _{n}$, in addition to the slower
oscillations we have just been considering. But,
except in a short initial period, the rapid oscillations disappear for all $%
n $ after smoothing with $H$.
\begin{figure}[htb]
\begin{center}
\scalebox{0.5}{\includegraphics{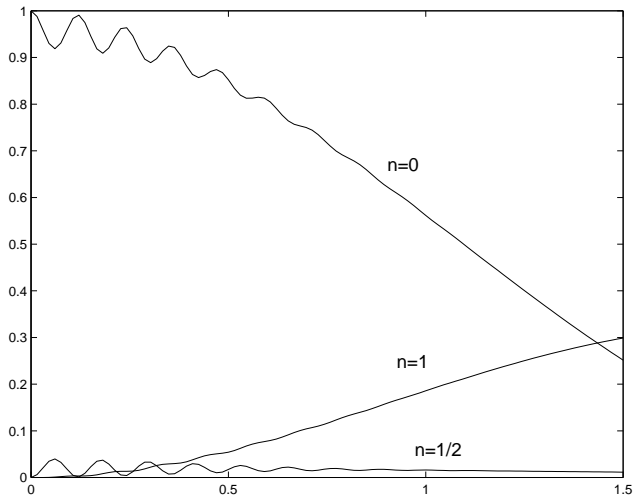}}
\end{center}
\caption{The intensities of the first few lines, averaged over the
velocity profile, in the QM model for small values of the transit
time $\tau$. The parameter $\gamma$ has been set at 0.2.}
\end{figure}
I have plotted the smoothed spectrum for the first few values of
$n$ in Fig.3; with $\gamma=0.2$ the rapid oscillations are
effectively damped out for $\tau
>1$. For the actual experimental range of $2<\tau <6,$ we may
smooth by putting simply $\left\langle 2\sin ^{2}\tau
_{n}\right\rangle =\left\langle 2\cos ^{2}\tau _{n}\right\rangle =1$ and $%
\left\langle \sin 2\tau _{n}\right\rangle =0$, leading to%
\begin{equation}
\overline{\rho _{n}^{\text{Q}}}\left( \tau \right)
=J_{n}^{2}\left( \tau \right) \left( 1-\gamma ^{2}\right) \text{ ,
}\overline{ \rho _{n+1/2}^{\text{Q}}}\left( \tau \right)
=\frac{\gamma ^{2}}{2}\left[ J_{n}^{2}\left( \tau \right)
+J_{n+1}^{2}\left( \tau \right) \right] \text{ ,}
\end{equation}%
and this corresponds to the characteristic function%
\begin{equation}
F_{\text{Q}}\left( \theta ;\tau \right) =F_{\text{Q}0}\left(
\theta ;\tau \right) \left[ 1+\gamma ^{2}\left( \cos \frac{\theta
}{2}-1\right) \right] \text{ .}  \label{charqm}
\end{equation}

\section{A stochastic model}

In his article, Gould\cite{gould} states that "\ldots if we
attempt to assign specific points in the diffraction pattern to
specific locations in the standing wave, we will fail miserably".
While not dissenting from this judgement, I stress that many areas
of classical physics produce situations of the same character; if
we were to try to predict the position of a Brownian particle
given its initial position and momentum, then we would fail
equally miserably. What probably motivated the statement is the
Heisenberg Inequality as applied to an atom of the beam. Since its
transverse momentum is defined by the collimation process to be a
small
fraction of $h/\lambda $, its position "uncertainty" is a large multiple of $%
\lambda $; this is reflected in our choice of initial wave
function $\psi \left( 0;\zeta \right) =1$ in the previous section,
giving uniform probability for all $\zeta $. However, the maximum
deflection, in a typical case where four even lines are visible
either side of the central line and
the transverse momentum at entrance to the beam is zero, is 2.4.10$^{-4}$%
rad, so, for a laser of width 0.1mm, the maximum change in the value of $\zeta $ at exit is 24nm or 0.04$%
\lambda $. Although the observed variable is the momentum, which means that $%
\zeta $, in QM parlance, is a "hidden" variable, I assert that it
is not unreasonable to maintain that, to within the atomic
diameter, somewhat less than 1nm, each atom in the beam has a well
defined $\zeta $, which varies only slightly as the atom crosses
the laser. For the moment I shall confine the model to the case
$\gamma \rightarrow 0,$ which means we are assuming the upper
internal state of the atom is infinitely short lived and only even
lines of the momentum spectrum are seen.

The model I propose is that of a Markov process on the set of integers $%
n\left( \tau ;\zeta \right) $, the transition matrix being
$\lambda _{mn}\left( \zeta \right) $, that is the probability of a
transition from $n$ to $m$ in an interval $\delta \tau $ is
$\lambda _{mn}\left( \zeta \right) \delta \tau +o\left( \delta
\tau \right) $. The Markov property means we assume that
transitions in successive intervals occur independently. I shall
make two further assumptions: (\emph{i}) the process is single-step, so $%
\lambda _{mn}=0$ unless $m=n\pm 1;$ this is suggested by the property (\ref%
{sstp}) of the QM process (\emph{ii}) the process is homogeneous,
and therefore\cite{bart} additive, so $\lambda _{m+r,n+r}=\lambda
_{mn}$. With these assumptions, the transition matrix has only two
independent
components, denoted $\lambda _{n+1,n}=\alpha \left( \zeta \right) $ and $%
\lambda _{n-1,n}=\beta \left( \zeta \right) $.

To summarize, the stochastic model of the process associates, with
a specific location $\zeta $ in the standing wave, a specific
Markov process with the parameters $\alpha \left( \zeta \right)
,\beta \left( \zeta \right) $.

The characteristic function for $n\left( \tau ;\zeta \right) $ is%
\begin{equation}
f_{\text{S}}\left( \theta ;\tau ;\zeta \right) =\left\langle
e^{in\theta }\right\rangle =\exp \left[ \tau \alpha \left( \zeta
\right) \left( e^{i\theta }-1\right) +\tau \beta \left( \zeta
\right) \left( e^{-i\theta }-1\right) \right] \text{ ,}
\label{charstoc}
\end{equation}%
and its occupation probabilities are%
\begin{equation}
P_{n}\left( \tau ;\zeta \right) =\left( \frac{\alpha }{\beta
}\right)
^{n/2}\exp \left[ -\alpha \tau -\beta \tau \right] I_{n}\left( 2\tau \sqrt{%
\alpha \beta }\right) \text{ .}
\end{equation}%
The predicted line intensities are obtained by integrating over
the "hidden"
variable $\zeta $, that is%
\begin{equation}
\rho _{n}^{\text{S}}\left( \tau \right) =\frac{1}{\pi }\int_{-\pi
/2}^{\pi /2}P_{n}\left( \tau ;\zeta \right) d\zeta \text{ ,}
\label{specstoc}
\end{equation}%
where $P_{n}\left( \tau ;\zeta \right) =P_{n}\left( \tau ;\zeta
+\pi \right) .$ I shall assume further that $\alpha \left( \zeta
+\pi /2\right) =\beta
\left( \zeta \right) $, which results in%
\begin{equation}
P_{n}\left( \tau ;\zeta \right) =P_{-n}\left( \tau ;\zeta +\pi
/2\right) \text{ ,}
\end{equation}%
and hence%
\begin{equation}
\rho _{n}^{\text{S}}\left( \tau \right) =\rho
_{-n}^{\text{S}}\left( \tau \right) =\frac{2}{\pi }\int_{0}^{\pi
/2}P_{n}\left( \tau ;\zeta \right) d\zeta \text{ .}
\end{equation}

A concrete example of such a model is%
\begin{equation}
\alpha \left( \zeta \right) =\frac{1+\sin 2\zeta }{2}\quad ,\quad
\beta \left( \zeta \right) =\frac{1-\sin 2\zeta }{2}\text{ .}
\label{concstoc}
\end{equation}%
This model is depicted in Fig.4,
\begin{figure}[htb]
\begin{center}
\scalebox{0.5}{\includegraphics{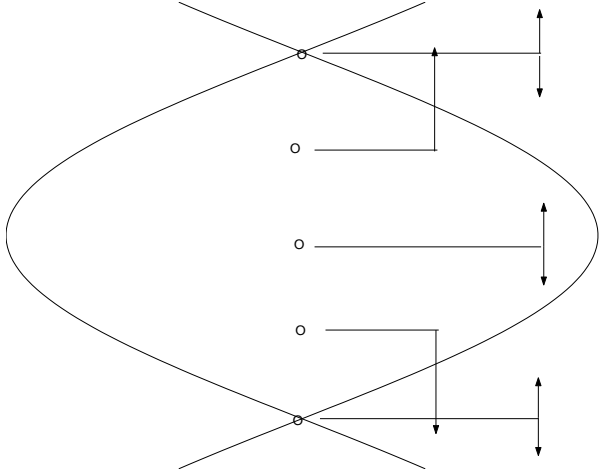}}
\end{center}
\caption{The Kapitza-Dirac effect according to the stochastic
model. For a given location of the atom in the stationary wave,
there is a pair of
transition probabilities, for  transverse impulses of $\pm 2h/\protect%
\lambda $ respectively. The sum of these probabilities is the same
for all locations. For example, the two probabilities are equal at
a node or an antinode, while one of them is zero at a point midway
between a node and an antinode; the atom moves preferentially
towards the nearest node.  }
\end{figure}
where the directions and magnitudes of the transition rates
$\alpha \left( \zeta \right) ,\beta \left( \zeta \right) $ are
indicated for a few different locations of the atom within the
standing
wave. Putting $\xi =\zeta -\pi /4$, the intensities become%
\begin{eqnarray}
\rho _{n}^{\text{S}}\left( \tau \right) &=&\frac{2e^{-\tau }}{\pi }%
\int_{0}^{\pi /2}\tan ^{n}\xi \text{ }I_{n}\left( \tau \sin 2\xi
\right) d\xi
\notag \\
&=&\frac{e^{-\tau }}{\pi }\sum_{r=0}^{\infty }\frac{\Gamma \left(
r+1/2\right) \Gamma \left( n+r+1/2\right) }{r!\left( n+r\right)
!\left( n+2r\right) !}\text{ }\tau ^{n+2r}\text{ ,}
\end{eqnarray}%
and the characteristic function is (compare eqn(7))%
\begin{equation}
F_{\text{S}}\left( \theta ;\tau \right) =\sum_{n=-\infty }^{\infty
}\rho _{n}^{\text{S}}\left( \tau \right) e^{in\theta }=\exp \left[
-\tau \left( 1-\cos \theta \right) \right] J_{0}\left( \tau \sin
\theta \right) \text{ .}
\end{equation}

\section{Comparison of the models}

The latter model has a variance%
\begin{equation}
\left\langle n^{2}\right\rangle _{\text{S}}=\frac{1}{2}\tau
^{2}+\tau \text{ ,}
\end{equation}%
as compared with the QM variance of $\tau ^{2}/2$. Whilst the
variances become indistinguishable for large $\tau $, for small
$\tau $ there is an essential difference; the initial variance is
of order $\tau ^{2}$ in the QM model, and of order $\tau $ in the
stochastic model. I postpone discussion of this disagreement to
the Discussion section.

In making a more detailed examination of the spectrum we begin by
comparing the asymptotics of the QM and stochastic models in the
limit $\tau \rightarrow +\infty .$ The QM intensities were
obtained in Section 2, and we now compare them with the
asymptotics of the stochastic model, which are
obtained from its characteristic function%
\begin{equation}
G_{\text{S}}\left( \tau ;z\right) =\frac{1}{\pi }\int_{-\pi /2}^{\pi /2}\exp %
\left[ \frac{1}{2}z\left( 1+\sin 2\zeta \right)
+\frac{1}{2}z^{-1}\left( 1-\sin 2\zeta \right) -1\right] d\zeta
\text{ ,}
\end{equation}%
namely%
\begin{equation}
\rho _{n}^{\text{S}}\left( \tau \right) =\frac{1}{2\pi i}\int_{C}G_{\text{S}%
}\left( \tau ;z\right) z^{-n-1}dz\text{ ,}
\end{equation}%
where $C$ is a closed contour enclosing the origin. This set of
functions has the surprisingly simple asymptotic behaviour
(obtained by selecting a
steepest-descent contour for $C$)%
\begin{equation}
\rho _{n}^{\text{S}}\left( \tau \right) \sim \frac{1}{\pi
\sqrt{\tau ^{2}-n^{2}}}\qquad \left( |n|<\tau \right) \text{ ,}
\end{equation}%
which, on averaging over the velocities of the atomic beam, gives \emph{%
exactly the same} asymptotics as the QM\ model, that is (see Fig.2)%
\begin{equation}
\overline{\rho _{n}^{\text{S}}}\left( \tau \right) =\int_{0}^{\infty }\rho _{n}^{\text{%
S}}\left( \tau ^{\prime }\right) H\left( \tau ,\tau ^{\prime
}\right) d\tau ^{\prime }\sim \overline{\rho
_{n}^{\text{Q}0}}\left( \tau \right) \text{ .}
\end{equation}

Now, turning to small values of $\tau $, I\ \ plot, in Fig.5,
\begin{figure}[htb]
\begin{center}
\scalebox{0.5}{\includegraphics{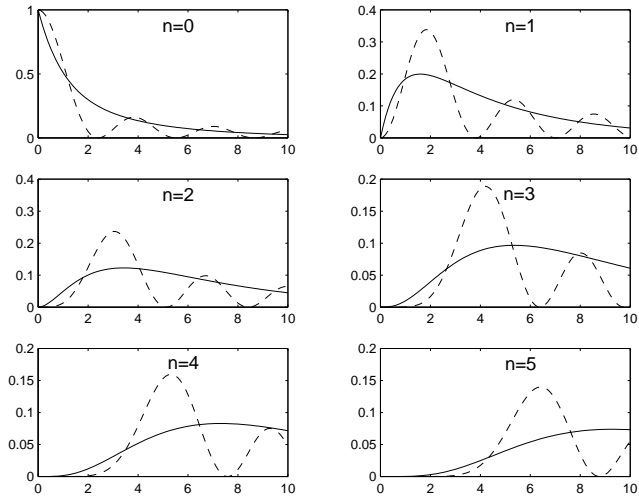}}
\end{center}
\caption{The  momentum spectrum of the Kapitza-Dirac effect
according to the stochastic model. The continuous lines represent
the intensity of the $n$th line as a function of the time
$\protect\tau $ spent in the laser, and the dashed line represents
the equivalent intensity in the QM model. At $\protect\tau =3$
only the lines $n\leq 4$ are visible.}
\end{figure}
the unsmoothed intensities, that is $\rho _{n}^{\text{Q}0}$ and
$\rho _{n}^{\text{S}},$ of the first six lines as functions of
$\tau $. Note that the QM and the stochastic models agree as to
their orders of magnitude; in particular the latter model gives
just four visible lines on either side of the centre line for the
case $\tau =3$. However, these curves show the disagreement of
variances noted above; it shows up as a zero slope at the origin for all $%
\rho _{n}^{\text{Q}0}$, compared with a negative slope for $\rho _{0}^{\text{%
S}}$ and a positive slope for $\rho _{1}^{\text{S}}$; the zero
slope for higher $\rho _{n}^{\text{S}}$ is a consequence of the
single-step assumption which we made in constructing the
stochastic model.

A more serious disagreement is the oscillatory dependence of $\rho _{n}$ on $%
\tau $ in the QM model. Indeed that model\ predicts zero intensity
for $\rho _{n}\left( \tau \right) $ whenever $J_{n}\left( \tau
\right) $ has a zero. An averaging over $\tau ,$ as above, will
give a smoothed intensity which never completely vanishes, but
nevertheless, for small $\tau $, the oscillations persist even
with such smoothing. On the other hand, in our concrete stochastic
model $\rho _{0}$ decreases monotonically, while the other $\rho
_{n}$ rise to a single maximum and then decrease monotonically,
that is there are no zeros. This behaviour is common to the whole
family of
Markov models, as may be seen by considering the derivative%
\begin{equation}
\frac{\partial }{\partial \tau }\left[ \tau ^{-n}e^{-\alpha \tau
-\beta \tau }I_{n}\left( 2\tau \sqrt{\alpha \beta }\right) \right]
=-\tau
^{-n}e^{-\alpha \tau -\beta \tau }\left[ \left( \alpha +\beta \right) I_{n}-2%
\sqrt{\alpha \beta }I_{n+1}\right] \text{ .}
\end{equation}%
The right hand side is negative for all positive $\alpha ,\beta
,\tau $ and all nonnegative $n$, and therefore, substituting in
(\ref{specstoc}),
\begin{subequations}
\begin{equation}
\frac{d}{d\tau }\left[ \tau ^{-n}\rho _{n}^{\text{S}}\left( \tau \right) %
\right] <0\qquad \qquad \left( n\geq 0,\tau >0\right) \text{ .}
\label{belltype}
\end{equation}%
Thus $\rho _{0}^{\text{S}}\left( \tau \right) $ certainly
decreases monotonically, as does also $\overline{\rho
_{0}^{\text{S}}}\left( \tau \right) $ but
not $\rho _{0}^{\text{Q}}\left( \tau \right) $ or $\overline{\rho _{0}^{\text{Q}}%
}\left( \tau \right) $. For $n>0$ the implication of the
inequality is
somewhat more complicated, but it is certainly not satisfied by $\rho _{n}^{%
\text{Q}}\left( \tau \right) $.

\section{Inclusion of odd-momentum states}

I shall now improve the model by including the odd states, so that
$n$ takes half-integral as well as integral values. A jump of $\pm
1/2$ from an even (that is integral $n$) state occurs with
probability $\delta \tau $, that is either direction is equally
probable, and a jump of $+1/2$ from an odd state occurs with
probability $\delta \tau \left( 1+\sin 2\zeta \right) /\gamma
^{2}$, while a jump of $-1/2$ from an odd state occurs with probability $%
\delta \tau \left( 1-\sin 2\zeta \right) /\gamma ^{2}$. Adopting
the standard classification of stochastic processes, the new model
may be
described as a pair of coupled additive processes, an additive process\cite%
{bart} being one which is homogeneous and Markov. Because of the factors of $%
\gamma ^{-2}$, this model gives a correction to $F_{\text{S}0}$
even to zero order, but we shall see that such corrections are
substantial only for $\tau $ of order $\gamma ^{2}$. Outside of this range, the corrections to $F_{%
\text{S}0}$ are of order $\gamma ^{2}$ only, so they do not
significantly reduce the disagreements we have just found between
$\rho _{n}^{\text{S}}$ and $\rho _{n}^{\text{Q}}$.

The characteristic function for the new model is
\end{subequations}
\begin{equation}
F_{\text{S}}\left( \theta ;\tau \right) =\frac{1}{\pi }\int_{-\pi
/2}^{\pi /2}f_{\text{S}}\left( \theta ;\tau ;\zeta \right) d\zeta
\text{ ,}
\end{equation}%
where the function $f_{\text{S}}\left( \theta ;\tau ;\zeta \right)
$ may be
decomposed into parts coming from even and odd states, that is%
\begin{equation}
f_{\text{S}}=f_{1}+f_{2}\text{ , } f_{1}=\sum_{-\infty }^{\infty
}P_{n}\left( \tau ;\zeta \right) e^{in\theta }\text{ , }
f_{2}=\sum_{-\infty }^{\infty }P_{n+1/2}\left( \tau ;\zeta \right)
e^{i\left( n+1/2\right) \theta }\text{ ,}
\end{equation}%
the $P_{n} $and $P_{n+1/2}$ being the occupation probabilities of
the even
and odd states. Then $f_{1},f_{2}$ have the time derivatives%
\begin{eqnarray}
\dot{f}_{1} &=&2\gamma ^{-2}\left[ \cos \left( \theta /2\right)
+i\sin
\left( \theta /2\right) \sin 2\zeta \right] f_{2}-2f_{1}\text{ ,}  \notag \\
\dot{f}_{2} &=&2\cos \left( \theta /2\right) f_{1}-2\gamma
^{-2}f_{2}
\end{eqnarray}%
with the initial values $f_{1}\left( \theta ;0;\zeta \right)
=1,f_{2}\left(
\theta ;0;\zeta \right) =0$. The solution is%
\begin{eqnarray}
f_{1} &=&\frac{\left( \gamma _{1}-2\right) e^{-\gamma _{2}\tau
}-\left( \gamma _{2}-2\right) e^{-\gamma _{1}\tau }}{\gamma
_{1}-\gamma _{2}}\text{ ,}
\notag \\
f_{2} &=&2\cos \left( \frac{\theta }{2}\right) \frac{e^{-\gamma
_{2}\tau }-e^{-\gamma _{1}\tau }}{\gamma _{1}-\gamma _{2}}\text{
,}  \label{sspec}
\end{eqnarray}%
where%
\begin{equation}
\gamma _{1,2}=\frac{1+\gamma ^{2}\pm \sqrt{1+2\gamma ^{2}\left(
\cos \theta +i\sin \theta \sin 2\zeta \right) +\gamma
^{4}}}{\gamma ^{2}}\text{ .}
\end{equation}%
To order $\gamma ^{2}$, and for $\gamma ^{2}\ll \tau \ll \gamma
^{-2}$, we
may discard the terms in $e^{-\gamma _{1}\tau }$ to obtain%
\begin{equation}
f_{\text{S}}=f_{1}+f_{2}=e^{-\gamma _{20}\tau }\left[ 1+\frac{\gamma ^{2}}{2}%
\left( 2\cos \frac{\theta }{2}-2+\left( 2\tau +1\right) \gamma
_{20}-\tau \gamma _{20}^{2}\right) \right]
\end{equation}%
where%
\begin{equation}
\gamma _{20}=1-\cos \theta -i\sin \theta \sin 2\zeta \text{ .}
\end{equation}%
Averaged over $\zeta $ this gives%
\begin{equation}
F_{\text{S}}\left( \theta ;\tau \right) =F_{\text{S}0}\left(
\theta ;\tau
\right) \left[ 1+\gamma ^{2}\left( \cos \frac{\theta }{2}-1\right) \right] -%
\frac{\gamma ^{2}\left( 2\tau +1\right)
}{2}\dot{F}_{\text{S}0}\left( \theta ;\tau \right) -\frac{\gamma
^{2}\tau }{2}\ddot{F}_{\text{S}0}\left( \theta ;\tau \right)
\end{equation}%
The spectrum, for $\gamma ^{2}\ll \tau \ll \gamma ^{-2}$, is then%
\begin{equation}
\rho _{n}=\rho _{n}^{0}\left( 1-\gamma ^{2}\right) -\frac{\gamma
^{2}\left(
2\tau +1\right) }{2}\dot{\rho}_{n}^{0}-\frac{\gamma ^{2}\tau }{2}\ddot{\rho}%
_{n}^{0}\text{ , } \rho _{n+1/2}=\frac{\gamma ^{2}}{2}\left( \rho
_{n}+\rho _{n+1}\right) \text{ ,}  \label{rhomark2}
\end{equation}%
and it may be extended to the range $\tau \ll \gamma ^{-2}$ by
adding
the effect of the terms in $e^{-\gamma _{1}\tau },$ that is%
\begin{equation}
\Delta \rho _{0}=\frac{\gamma ^{2}}{2}\text{ }e^{-2\tau
_{0}}\text{ },\text{
}\Delta \rho _{1/2}=-\frac{\gamma ^{2}}{2}\text{ }e^{-2\tau _{0}}\text{ },%
\text{ }\Delta \rho _{1}=\frac{\gamma ^{2}}{4}\text{ }e^{-2\tau
_{0}}\text{ , }\Delta \rho _{n}=0\text{ \ }\left( n>1\right)
\text{ ,}  \label{deltarho}
\end{equation}
the quantity $\tau _{0}$ having been defined in (\ref{taun}). In
Fig.6
\begin{figure}[htb]
\begin{center}
\scalebox{0.5}{\includegraphics{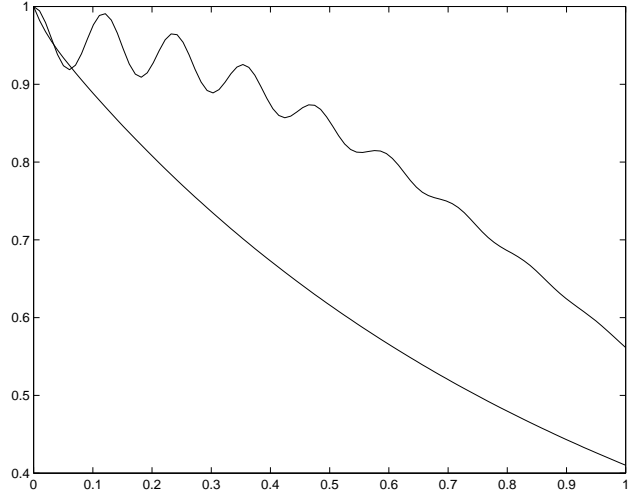}}
\end{center}
\caption{The intensity of the centre line as a function of the
transit time in the interval $0<\tau<1$. The upper curve depicts
the QM model, and the lower curve the stochastic model. The
parameter $\gamma$ has been taken as 0.2.}
\end{figure}
I have plotted $\rho _{0}^{\text{S}}\left( \tau \right) $ in the
range $0<\tau <1,$ including a plot of $\overline{\rho
_{0}^{\text{Q}}}\left( \tau \right) $ for comparison.

The new stochastic model indicates the role of the states of odd
momentum. They have an intensity of order $\gamma ^{2}$, because
the transition time from an upper to a lower atomic state is
smaller than that of the reverse transition by a factor of order
$\gamma ^{2}$. This indicates a crucial role for the  zeropoint
electromagnetic field (ZPEF) which has also played an
important role in the program, developed by Emilio Santos and myself\cite%
{bellcrit}\cite{bellcrit1}, and  designed to achieve a local
realist understanding of the optical Bell experiments. When the
detuning frequency exceeds the resonant Rabi frequency,
"spontaneous" transitions, that is transitions induced by the
ZPEF, are more frequent than laser-induced ones. Note that the new
stochastic model differs radically from the QM model for the case
that the incoming atom is in its upper state, because in that case
the fast transition, with probability proportional to $\gamma
^{-2}$, occurs first. This gives a spectrum with strong lines at
$n=\pm 1/2,\pm 3/2\ldots ,$ and weak lines at $n=0,\pm 1,\pm
2\ldots ,$ that is a reversal of the pattern shown for an incoming
atom in its lower state. The QM model predicts no difference
between these two spectra, so the discrepancy may provide an
experimental method for determining which is the more correct out
of the two models.

\section{Discussion}

Before discussing the disagreements between the QM and stochastic
models, I emphasize the agreement we obtained in the asymptotic
limit $\tau \rightarrow \infty .$ It is easily shown that the
choice of $\alpha \left( \zeta \right) $ and $\beta \left( \zeta
\right) $ made in (\ref{concstoc}) is, up to a phase shift in
$\zeta $, the only one which gives asymptotic agreement with the
QM model. There is a simple explanation for this, namely
that, in the stochastic model, the drift in an interval $\tau $ is $\tau %
\left[ \alpha \left( \zeta \right) -\beta \left( \zeta \right)
\right] $, which, with the choice we have made, becomes $\tau \sin
2\zeta $. This, without diffusion, is precisely the deterministic
model occurring in Section 2 as the classical limit of the QM
model (see Fig.2). Hence the deterministic parts of both the QM
and stochastic models give the same results.

The disagreement between the models, which we found for very small
$\tau $, may well be irrelevant, since the QM model has very rapid
oscillations (see Fig.3), and we have just seen that the modified
Markov model also produces\ a radical change in the intensities
for very small $\tau $. The quantum mechanical behaviour, whereby
the initial probability of transition from a pure state changes
from 1 by a quantity proportional to $\tau ^{2},$ is a general
characteristic, called the Quantum Zeno Paradox, according to
which a continuously observed system cannot undergo a transition.
This paradox has never received a satisfactory resolution. On the
other hand $\left\langle n^{2}\right\rangle $ is proportional
initially to $\tau $ for any Markov process. Note, however, that,
although Fig.6, like the first diagram in Fig.5, does indeed show
an initial decrease in $\rho _{0}^{\text{S}}$
proportional to $\tau $, compared with $\tau ^{2}$ for both $\rho _{0}^{%
\text{Q}0}$ and $\overline{\rho _{0}^{\text{Q}}},$ the initial
curvature of the latter is very large compared with that of the
former$,$ which means that direct observation of the Zeno
phenomenon would be extremely difficult.

I turn finally to the disagreement shown in Fig.5, in particular
the oscillatory behaviour of $\rho _{n}^{\text{Q}0}\left( \tau
\right) $. We need to know the extent to which this behaviour of
the line intensity is actually supported by experiment, in
particular whether the observed spectrum is consistent with
(\ref{belltype}). If the existence of zero-intensity lines for
certain $\tau $ is confirmed by experimental evidence, then the
class of stochastic processes may have to be extended to allow for
the possibility that the atom has a memory of a recent transition
having occurred.

The comparison I have made between the QM and stochastic models,
or between the wavelike "atom" of Fig.1 and the more recognizably
atomic object of Fig.4, indicates to me that the atom of
Democritus, or of Boltzmann, is by no means dead. Inequality
(\ref{belltype}) provides us with a means to determine which of
these simple models gives the better agreement with experiment. It
would also be interesting to try repeating the experiment with an
incoming beam of atoms in the upper state, to see whether the
pattern of strong and weak lines is actually inverted, as
predicted in the stochastic model.

\end{document}